\documentclass[11pt]{article}
\usepackage{epsfig}
\usepackage{rotating}

\def\OEE{\Omega_{\rm IB}}

\newcommand{\RE}{{\rm Re}}
\newcommand{\IM}{{\rm Im}}
\newcommand{\vcb}{|V_{cb}|}
\newcommand{\vtd}{|V_{td}|}
\newcommand{\vub}{|V_{ub}/V_{cb}|}

\def\ps{\rm ps}
\def\R1{\varepsilon_1}
\def\E8{\varepsilon_8}

\def\epe{\varepsilon'/\varepsilon}

\newcommand{\mt}{m_{\rm t}}
\newcommand{\mtb}{\overline{m}_{\rm t}}

\newcommand{\mc}{m_{\rm c}}
\newcommand{\ms}{m_{\rm s}}

\newcommand{\mb}{m_{\rm b}}
\newcommand{\mw}{M_{\rm W}}
\newcommand{\mz}{M_{\rm Z}}
\newcommand{\gev}{\, {\rm GeV}}
\newcommand{\mev}{\, {\rm MeV}}
\newcommand{\bsi}{B_6^{(1/2)}}
\newcommand{\bei}{B_8^{(3/2)}}
\newcommand{\Lms}{\Lambda_{\overline{\rm MS}}}

\newcommand{\bea}{\begin{eqnarray}}
\newcommand{\eea}{\end{eqnarray}}
\newcommand{\bd}{\begin{displaymath}}
\newcommand{\ed}{\end{displaymath}}

\newcommand{\beq}{\begin{equation}}
\newcommand{\eeq}{\end{equation}}
\newcommand{\be}{\begin{equation}}
\newcommand{\ee}{\end{equation}}
\newcommand{\bi}{\begin{itemize}}
\newcommand{\ei}{\end{itemize}}
\newcommand{\ord}{{\cal O}}

\def\kpn{K^+\rightarrow\pi^+\nu\bar\nu}
\def\klpn{K_{\rm L}\rightarrow\pi^0\nu\bar\nu}
\newcommand{\kmm}{K_{\rm L} \to \mu^+ \mu^-}
\newcommand{\kpe}{K_{\rm L} \to \pi^0 e^+ e^-}

\newcommand{\imlt}{\IM\lambda_t}

\begin{document}
\thispagestyle{empty}
\phantom{xxx}
\vskip1truecm
\begin{flushright}
 TUM-HEP-435/01 \\
September 2001
\end{flushright}
\vskip1.8truecm
\centerline{\LARGE\bf Flavour Physics and 
 CP Violation in the SM }
   \vskip1truecm
\centerline{\Large\bf Andrzej J. Buras}
\bigskip
\centerline{\sl Technische Universit{\"a}t M{\"u}nchen}
\centerline{\sl Physik Department} 
\centerline{\sl D-85748 Garching, Germany}
\vskip1truecm
\centerline{\bf Abstract}
We review the main aspects of Flavour Physics and CP Violation in the 
Standard Model. After presenting a grand view of the field
including a Master Formula for weak decays we discuss
i) Standard analysis of the unitarity triangle, 
ii)  The ratio $\epe$,
iii) Rare decays $K^+\to\pi^+\nu\bar\nu$ and $K_L\to\pi^0\nu\bar\nu$,
iv) CP violation in B decays and v) Models with Minimal Flavour
Violation. Our review ends with 20 questions that hopefully will be 
answered in the coming years.
\vskip1.6truecm

\centerline{\it Introductory Lecture given at }
\centerline{\bf KAON 2001}
\centerline{\it Pisa, 12 June--17 June, 2001}

\newpage

\section{Introduction}

The field of Flavour Physics and CP Violation constitutes an important 
part of the Standard Model (SM). It will certainly be one of the 
hot topics in particle physics during this decade. In this introductory 
lecture I will attempt to describe this field in general terms paying special 
attention to the theoretical framework and to a few selected topics which 
in my opinion are very important. Instead of an outlook I will provide a 
list of twenty questions for KAON 2001 and beyond, that will allow me to 
address other important topics. In view of considerable space 
limitations it is impossible to refer properly to the relevant literature. 
As a compensation, references to roughly 800 papers can be found  
in my Erice lectures \cite{Erice}. 

\section{Grand View}
\setcounter{equation}{0}

There are four basic properties in the SM that govern flavour physics and 
CP violation in this model. These are 
\begin{itemize}
\item
Breakdown of Parity: 
charged current interactions are only between left-handed 
quarks and between left-handed leptons.
\item
Quark Mixing: 
 the {\it weak
eigenstates} $(d^\prime,s^\prime,b^\prime)$ of quarks differ from 
 the corresponding {\it mass eigenstates} $d,s,b$:
\begin{equation}\label{2.67}
\left(\begin{array}{c}
d^\prime \\ s^\prime \\ b^\prime
\end{array}\right)=
\left(\begin{array}{ccc}
V_{ud}&V_{us}&V_{ub}\\
V_{cd}&V_{cs}&V_{cb}\\
V_{td}&V_{ts}&V_{tb}
\end{array}\right)
\left(\begin{array}{c}
d \\ s \\ b
\end{array}\right)\equiv\hat V_{\rm CKM}\left(\begin{array}{c}
d \\ s \\ b
\end{array}\right).
\end{equation}
The unitary transformation connecting these states is the 
Cabibbo-Kobayashi-Maskawa (CKM) matrix.
\item 
GIM Mechanism: 
The unitarity of the CKM-matrix assures the absence of 
flavour changing neutral current transitions at the tree level. 
These processes can consequently appear first at the one-loop 
level and are very sensitive to short distance flavour 
dynamics.
\item
Asymptotic Freedom in QCD: 
Whereas strong interaction effects at short distance scales 
$\mu_{SD}=\ord(\mw,\mz,\mt)$ can be treated by perturbative methods, 
at long distance scales, $\mu_{LD}=\ord(1-2\,\gev)$, the use of 
non-perturbative methods 
becomes mandatory. The latter fact brings considerable 
uncertainties in the theoretical predictions. The appearance 
of two vastly different scales implies large $\log \mu_{SD}/\mu_{LD}$ 
multiplying $\alpha_s$
 that fortunately can be summed up to all orders of 
perturbation theory in $\alpha_s$ by means of renormalization group 
methods.
\end{itemize}

\begin{figure}[hbt]
\vspace{0.10in}
\centerline{
\epsfysize=2.1in
\epsffile{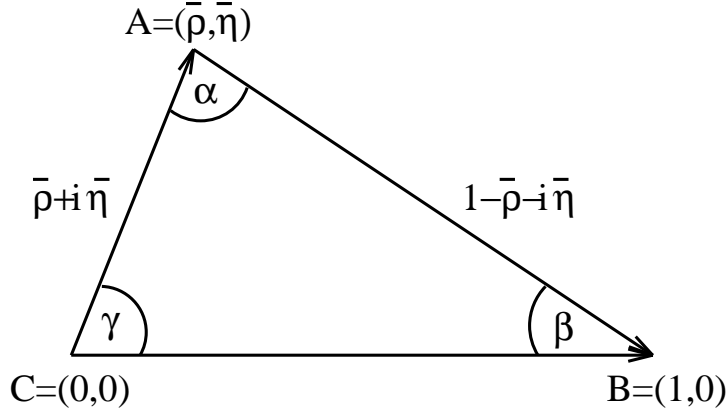}
}
\vspace{0.08in}
\caption{Unitarity Triangle.}\label{fig:utriangle}
\end{figure}

According to the Kobayashi-Maskawa picture of CP violation, this 
phenomenon arises from a single complex phase $\delta_{KM}$ in the 
$W^\pm$--interactions of quarks. The CKM matrix can be parametrized by three 
mixing angles and $\delta_{KM}$ as described by the standard parametrization 
that is recommended by the Particle Data Group. While this standard 
parametrization should certainly be recommended for calculations, in a 
talk like this one, the Wolfenstein parametrization \cite{WO} 
is certainly more useful: 
\begin{equation}\label{2.75} 
\hat V_{\rm CKM}=
\left(\begin{array}{ccc}
1-{\lambda^2\over 2}&\lambda&A\lambda^3(\varrho-i\eta)\\ -\lambda&
1-{\lambda^2\over 2}&A\lambda^2\\ A\lambda^3(1-\varrho-i\eta)&-A\lambda^2&
1\end{array}\right)
+\ord(\lambda^4)\,.
\end{equation}
Including the most important $\ord(\lambda^4)$ and higher order terms
one finds then that to an excellent accuracy
\begin{equation}\label{CKM1}
V_{us}=\lambda, \qquad V_{cb}=A\lambda^2, \qquad V_{ts}=-A\lambda^2,
\end{equation}
\begin{equation}\label{CKM2}
V_{ub}=A\lambda^3(\varrho-i\eta),
\qquad
V_{td}=A\lambda^3(1-\bar\varrho-i\bar\eta)
\end{equation}
where $\lambda, A, \varrho, \eta$ are the Wolfenstein parameters and
\begin{equation}\label{2.88d}
\bar\varrho=\varrho (1-\frac{\lambda^2}{2}),
\qquad
\bar\eta=\eta (1-\frac{\lambda^2}{2})
\end{equation}
are the parameters introduced in \cite{BLO}. 
The latter parameters describe the 
apex of the unitarity triangle (UT) shown in fig.~\ref{fig:utriangle}
 with the length CA, BA, 
and 
CB equal respectively to
\begin{equation}\label{2.94}
R_b= (1-\frac{\lambda^2}{2})\frac{1}{\lambda}
\left| \frac{V_{ub}}{V_{cb}} \right|,\qquad
R_t =\frac{1}{\lambda} \left| \frac{V_{td}}{V_{cb}} \right|,
\qquad 1~.
\end{equation}
The angles $\beta$ and $\gamma$ of the UT determine 
the complex phases of the CKM-elements $V_{td}$ and
$V_{ub}$:
\be\label{e417}
V_{td}=|V_{td}|e^{-i\beta},\qquad V_{ub}=|V_{ub}|e^{-i\gamma}.
\ee

\begin{figure}[hbt]
  \vspace{0.10in} \centerline{
\begin{turn}{-90}
  \mbox{\epsfig{file=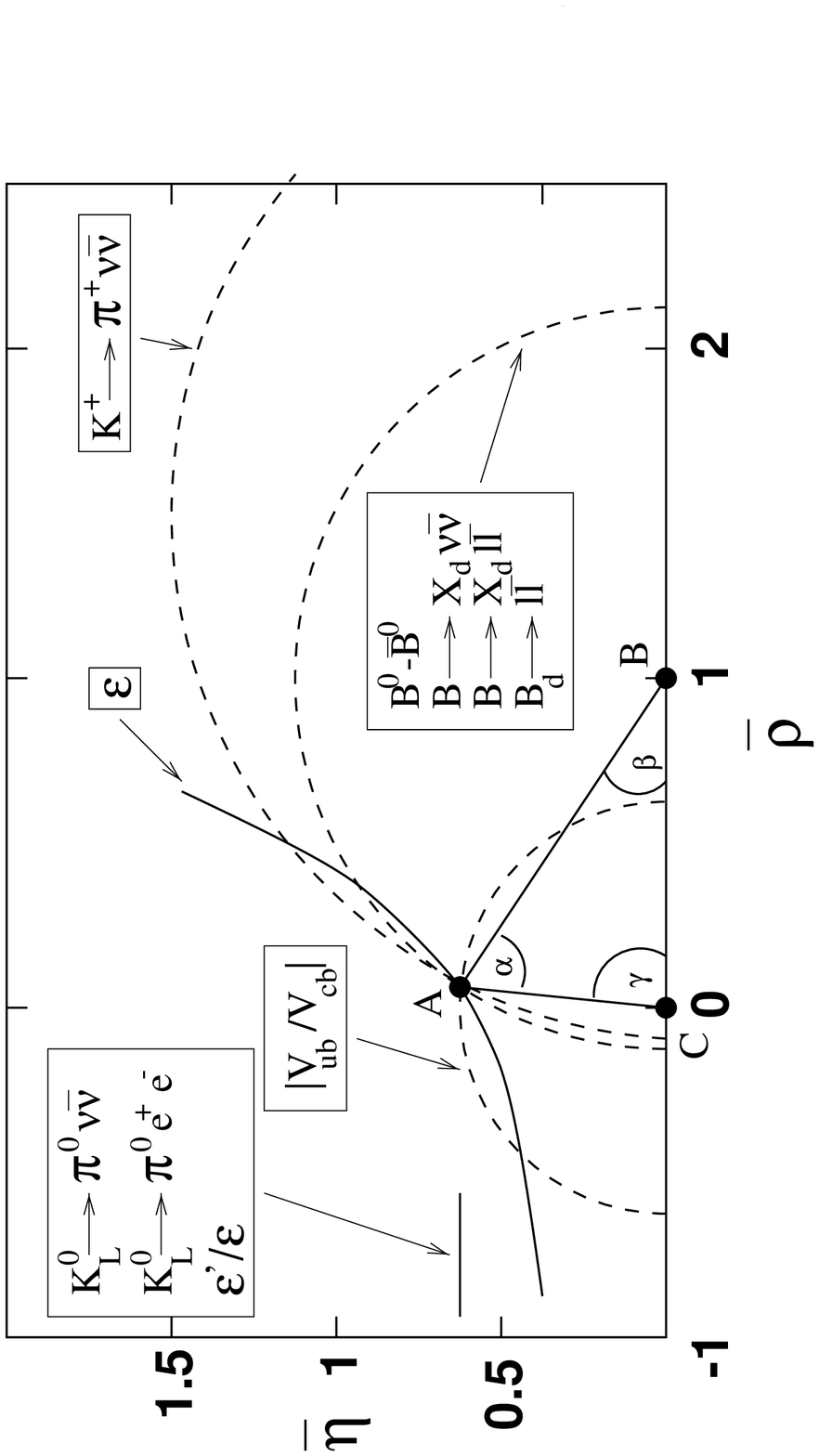,width=0.4\linewidth}}
\end{turn}
} \vspace{0.08in}
\caption[]{
  The ideal Unitarity Triangle.
\label{fig:2011}}
\end{figure}

The apex $(\bar\varrho,\bar\eta)$ of the UT can be efficiently hunted by 
means of 
rare and CP violating transitions as shown in fig.~\ref{fig:2011}. Moreover 
the angles 
of this triangle can be measured in CP asymmetries in B-decays and 
using other strategies. This picture could describe in principle the 
reality in the year 2011, my retirement year, if the SM is the whole story. 
On the other hand in the 
presence of significant new physics contributions, the use of the SM 
expressions for rare and CP violating transitions in question, combined 
with future precise measurements, may result in curves which do not cross 
each other at a single point in the $(\bar\varrho,\bar\eta)$ plane. 
This would be truly exciting and most of us hope that this will turn out 
to be the case. In order to be able to draw such thin curves as in 
fig.~\ref{fig:2011}, not only experiments but also the theory has to be 
under control. Let me then briefly discuss the theoretical framework for weak 
decays.

\section{Master Formula for Weak Decays} 
\setcounter{equation}{0}
The present framework for weak decays is based on the operator product 
expansion (OPE) that allows to separate short and long distance 
contributions to weak amplitudes and on the renormalization group (RG) 
methods that allow to sum large logarithms $\log \mu_{SD}/\mu_{LD}$  to 
all orders in perturbation theory. The full exposition of these methods 
can be found in \cite{AJBLH,BBL}. 
Here I just want to propose a master formula for weak decay amplitudes that 
follows from OPE and RG and goes beyond the SM. 
It reads:
\be\label{master}
{\rm A(Decay)}= \sum_i B_i \eta^i_{\rm QCD}V^i_{\rm CKM} 
\lbrack F^i_{\rm SM}+F^i_{\rm New}\rbrack +
\sum_k B_k^{\rm New} \lbrack\eta^k_{\rm QCD}\rbrack^{\rm New} V^k_{\rm New} 
\lbrack G^k_{\rm New}\rbrack\, .
\ee
The non-perturbative parameters $B_i$ represent the matrix elements of local 
operators present in the SM. For instance in the case of 
$K^0-\bar K^0$ mixing, the matrix element of the operator
$\bar s \gamma_\mu(1-\gamma_5) d \otimes \bar s \gamma^\mu(1-\gamma_5) d $
is represented by the parameter $\hat B_K$.
There are other non-perturbative parameters in the SM that represent 
matrix elements of operators $Q_i$ with different colour and Dirac 
structures. The objects $\eta^i_{\rm QCD}$ are the QCD factors resulting 
from RG-analysis of the corresponding operators and $F^i_{\rm SM}$ stand for 
the so-called Inami-Lim functions \cite{IL} that result from the calculations 
of various
box and penguin diagrams. They depend on the top-quark mass. 
$V^i_{\rm CKM}$ are 
the CKM-factors we want to determine. 
 
New physics can contribute to our master formula in two ways. It can 
modify the importance of a given operator, present already in the SM, 
through the new short distance functions $F^i_{\rm New}$ that depend on 
the new 
parameters in the extensions of the SM like the masses of charginos, 
squarks, charged Higgs particles and $\tan\beta=v_2/v_1$ in the MSSM. 
These new 
particles enter the new box and penguin diagrams. In more complicated 
extensions of the SM new operators (Dirac structures) that are either 
absent or very strongly suppressed in the SM, can become important. 
Their contributions are described by the second sum in 
(\ref{master}) with 
$B_k^{\rm New}, \lbrack\eta^k_{\rm QCD}\rbrack^{\rm New}, V^k_{\rm New}, 
G^k_{\rm New}$
being analogs of the corresponding objects in the first sum of the master 
formula. The $V^k_{\rm New}$ show explicitly that the second sum describes 
generally new sources of flavour and CP violation beyond the CKM matrix. 
This sum may, however, also include contributions governed by the CKM 
matrix that are strongly suppressed in the SM but become important in 
some extensions of the SM. A typical example is the enhancement of the 
operators with Dirac structures $(V-A)\otimes(V+A)$, 
$(S-P)\otimes (S\pm P)$ and 
$\sigma_{\mu\nu} (S-P) \otimes \sigma^{\mu\nu} (S-P)$ contributing to 
$K^0-\bar K^0$ and $B^0-\bar B^0$ mixings in the MSSM with large 
$\tan\beta $. The most recent compilation of references to existing NLO 
calculations of $\eta^i_{\rm QCD}$ and 
$\lbrack\eta^k_{\rm QCD}\rbrack^{\rm New}$ can be found in \cite{Erice}.
 
Clearly the new functions $F^i_{\rm New}$ and $G^k_{\rm New}$ as well as the 
factors $V^k_{\rm New}$ may depend on new CP violating phases complicating 
considerably phenomenological analysis. We will see this in Masiero's 
lecture. In the present talk, that is dominantly devoted to the SM, I 
will only consider the simplest class of the extensions of the SM 
in which the second sum in (\ref{master}) is absent (no new operators) 
and flavour changing transitions are 
governed by the CKM matrix. In particular there are no new complex 
phases beyond the CKM phase. I will call this scenario ``Minimal Flavour 
Violation" (MFV) \cite{CDGG,UUT} being aware of the fact that for some authors 
MFV means a more general framework in which also new operators can give 
significant contributions. In the MFV models, as defined in 
\cite{CDGG,UUT},  our master formula simplifies to 
\be\label{mmaster}
{\rm A(Decay)}= \sum_i B_i \eta^i_{\rm QCD}V^i_{\rm CKM} 
\lbrack F^i_{\rm SM}+F^i_{\rm New}\rbrack 
\ee 
with $F^i_{\rm SM}$ and $F^i_{\rm New}$ being real. 
\section{Five Topics} 
\setcounter{equation}{0}
\subsection{Standard Analysis of the Unitarity Triangle} 
This analysis uses $\lambda=|V_{us}|=0.222\pm0.002$,
\be\label{tree}
\vcb=0.041\pm0.002, \qquad 
\frac{|V_{ub}|}{\vcb}=0.085\pm0.018
\ee
and the following three constraints: 
\begin{itemize}
\item 
$\varepsilon_K$--Hyperbola (Indirect CP Violation in $K_L\to \pi\pi$):
\begin{equation}\label{100a}
\bar\eta \left[(1-\bar\varrho) A^2 \eta_{\rm QCD}^{tt} F_{tt}
+ P_c(\varepsilon) \right] A^2 \hat B_K = 0.204~,
\end{equation}
where $\eta_{\rm QCD}^{tt}=0.57\pm0.01$, 
$P_c(\varepsilon)=0.30\pm0.05$ represents charm contribution and
$F_{tt}=2.38\pm0.11$ is the Inami-Lim $(t,t)$ box diagram function, 
denoted often by $S_0(x_t)$.
\item
$B^0_d-\bar B^0_d$--Mixing Constraint: 
\begin{equation}\label{RT}
R_t= 0.85~ \left[\frac{0.83}{A}\right]\sqrt{\frac{2.38}{F_{tt}}}
\sqrt{\frac{\Delta M_d}{0.487/{\rm ps}}}
          \left[\frac{230~\mev}{\sqrt{\hat B_d} F_{B_d}}\right]
          \sqrt{\frac{0.55}{\eta_B^{\rm QCD}}}
\ee
where $A=0.83\pm0.04$, $\Delta M_d=(0.487\pm0.009)/\ps$ and 
$\eta_B^{\rm QCD}=0.55\pm0.01$.
\item 
$B^0_s-\bar B^0_s$--Mixing Constraint ($\Delta M_d/\Delta M_s$): 
\be\label{Rt}
R_t=0.94~\sqrt{\frac{\Delta M_d}{0.487/\ps}}
\sqrt{\frac{15.0/ps}{\Delta M_s}}\left[\frac{\xi}{1.15}\right],
\qquad \xi=\frac{\sqrt{\hat B_s}F_{B_s}}{\sqrt{\hat B_d}F_{B_d}}
\ee
where $\Delta M_s>15.0/\ps$ from LEP experiments. 
\end{itemize}
The main uncertainties in this analysis originate in the theoretical 
uncertainties in the parameters $\hat B_K$ and $\sqrt{\hat B_d}F_{B_d}$
and to a lesser extent in $\xi$: 
\be
\hat B_K=0.85\pm0.15, \qquad  \sqrt{\hat B_d}F_{B_d}=(230\pm40)~MeV,
\qquad \xi=1.15\pm0.06~.
\ee
Also the uncertainties in (\ref{tree}), in particular in 
$\vub$, are substantial. Reviews of lattice results for the parameters in 
question can be found in \cite{LATT}. 

\begin{figure}[thb]
\vspace{0.10in}
\centerline{
\epsfysize=2.8in
\epsffile{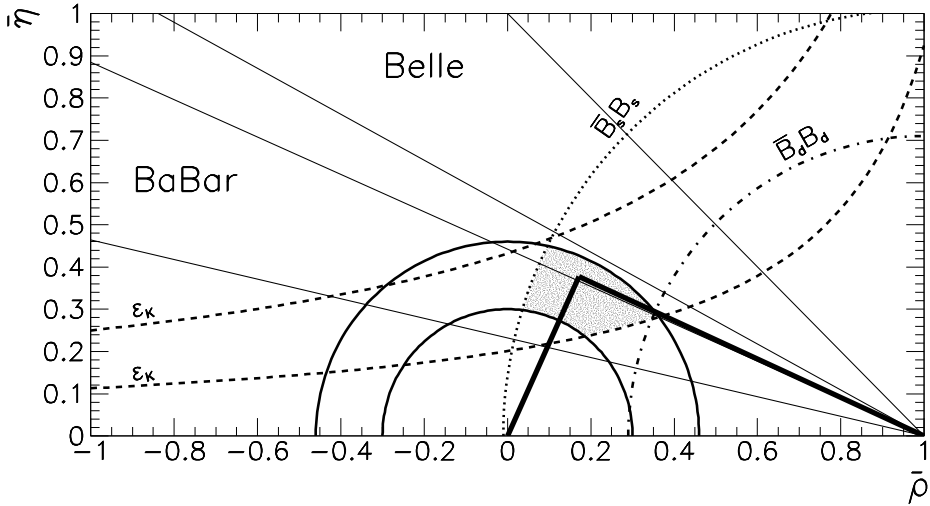}
}
\vspace{0.08in}
\caption[]{
Conservative Unitarity Triangle as of September 2001.
\label{fig:utdata}}
\end{figure}

One of the important issues is the error analysis of these formulae. In the 
literature five different methods are used: Gaussian approach \cite{Gaus}, 
Bayesian approach \cite{C00}, frequentist approach \cite{FREQ}, 
$95\%$ C.L. scan method \cite{SCAN95} and the simple 
(naive) scanning within one standard deviation as used 
by myself. Interestingly, the last method gives 
ranges for the output quantities that are similar to the $95\%$ C.L. 
ranges obtained by the remaining methods. To this end 
the same input parameters have to be used and the implementation of the 
lower bound on $\Delta M_s$ has to be  done in the same manner. 
Mele discusses these issues in his contibution. 
On my part I show in fig.~\ref{fig:utdata} the result of my own analysis 
that uses naive scanning. 
The allowed region for $(\bar\varrho,\bar\eta)$ 
is the shaded area on the right hand side of the  circle
representing  
the lower bound for $\Delta M_s$, that is $\Delta M_s>15/\ps$.
The hyperbolas in fig.~\ref{fig:utdata}
give the constraint from $\varepsilon$ and the two circles centered
at $(0,0)$ the constraint from $\vub$.
The circle on the right comes from $B^0_{d}-\bar B^0_{d}$ mixing
and excludes the region to its right. 
We observe that the region
$\bar\varrho<0$ is practically excluded by the lower bound on
$\Delta M_s$.
It is clear
from this figure that $\Delta M_s$ is a very important
ingredient in this analysis and that the measurement of $\Delta M_s$
giving also  lower bound on $R_t$ will have a large impact
on the plot in fig.~\ref{fig:utdata}. 

\begin{table}[hbt]
\caption[]{Output of the Standard Analysis. 
 $\lambda_t=V^*_{ts} V_{td}$.\label{TAB2}}
\vspace{0.4cm}
\begin{center}
\begin{tabular}{|c||c||c||c|}\hline
{\bf Quantity} & {\bf Scanning} & {\bf Bayesian I} & {\bf Bayesian II} 
 \\ \hline
$\bar\varrho$ & $ 0.07 - 0.34 $ & $0.14-0.30$ & $0.13-0.34$ \\ \hline
$\bar\eta$ & $0.22-0.45$ & $0.24-0.40$ & $0.22-0.46$ \\ \hline
$\sin(2\beta)$ &$0.50 - 0.84$ & $ 0.56-0.82$ & $0.52-0.92$ \\ \hline
$\sin(2\alpha)$ &$-0.87 - 0.36$ &$-0.83-0.04$ & $-0.85-0.14$ \\ \hline
$\gamma$ & $37.7^\circ - 75.7^\circ$ &$42.8^\circ-67.4^\circ$ &
$41.8^\circ-67.6^\circ$ \\ \hline
$\IM \lambda_t/10^{-4}$ &$0.94 - 1.60 $ &$ 0.93-1.43$ & $0.91-1.55$ \\ \hline
$\mid V_{td}\mid/10^{-3}$ &$6.7 - 9.3$ &$7.0-8.6$ & $6.8-8.7$ \\ \hline
\end{tabular}
\end{center}
\end{table}

The ranges for various quantities found using the scanning method are 
compared in table 1 with the $95\%$ C.L. ranges from Bayesian I of 
Ciuchini et al \cite{C00} that uses 
$\vcb=0.0410\pm0.0016$ and $\vub=0.086\pm 0.009$ 
and Bayesian II with $\vcb$ and $\vub$ as in (\ref{tree}) that are
used in my analysis. I thank Stocchi for providing the latter numbers.
My ranges are substantially larger than Bayesian I but only slightly 
larger than Bayesian II. This is partly related to a different treatment 
of the bound on $\Delta M_s$  done by Ciuchini et al and myself.
My ranges are very close to the ones obtained
using the frequentist approach \cite{FREQ}.

One of the highlights of this year were the improved measurements of 
$\sin2\beta$ by means of the time-dependent CP asymmetry
in $B^0_d(\bar B^0_d)\to \psi K_S$ decays 
\be
a_{\psi K_S}(t)\equiv -a_{\psi K_S}\sin(\Delta M_s t)=
-\sin 2 \beta \sin(\Delta M_s t)
\ee
with the last relation valid in those MFV models in which as in the SM 
$F_{tt}>0$ \cite{BF01}. The most recent measurements of $a_{\psi K_S}$ from
the BaBar and Belle Collaborations read
\begin{equation}\label{sinb}
(\sin 2\beta)_{\psi K_S} =\left\{ \begin{array}{ll}
0.59 \pm 0.14\pm 0.05 & {\rm (BaBar)}~\cite{BaBar} \\
0.99 \pm 0.14 \pm 0.06 &{\rm (Belle)}~ \cite{Belle}
\end{array} \right.
\end{equation}
and establish confidently CP violation in the B system!
A mile stone in the field of CP violation.
Combining these results with earlier measurements by CDF 
$(0.79^{+0.41}_{-0.44})$ and ALEPH $(0.84^{+0.82}_{-1.04}\pm 0.16)$ 
gives the grand average 
\be
a_{\psi K_S}=0.79\pm 0.10~.
\label{ga}
\ee
In view of the fact that the BaBar and Belle results are not fully consistent
with each other, the averaging of these results and the grand average 
given above could be questioned. Probably a better description of the 
present situation is $a_{\psi K_S}=0.80\pm 0.20~.$

In any case, these first direct measurements of the angle $\beta$ are
in a good agreement (see fig.~\ref{fig:utdata}) 
with the results of the standard analyses of
the unitarity triangle within the SM, even if the Belle result appears 
a bit too high. 
Clearly in view of a considerable difference between BaBar and Belle results
and still sizable uncertainty in the error estimates
of $(\sin 2\beta)_{\rm SM}$, there is a room for new physics 
contributions but the agreement of the prediction for $\sin 2\beta$ from
the fits of the unitarity triangle with the measured value 
of $a_{\psi K_S}$ is a strong indication that the CKM matrix could turn 
out to be the dominant source of CP violation in flavour violating decays.

\begin{figure}[thb]
\vspace{0.10in}
\centerline{
\epsfysize=4.1in
\epsffile{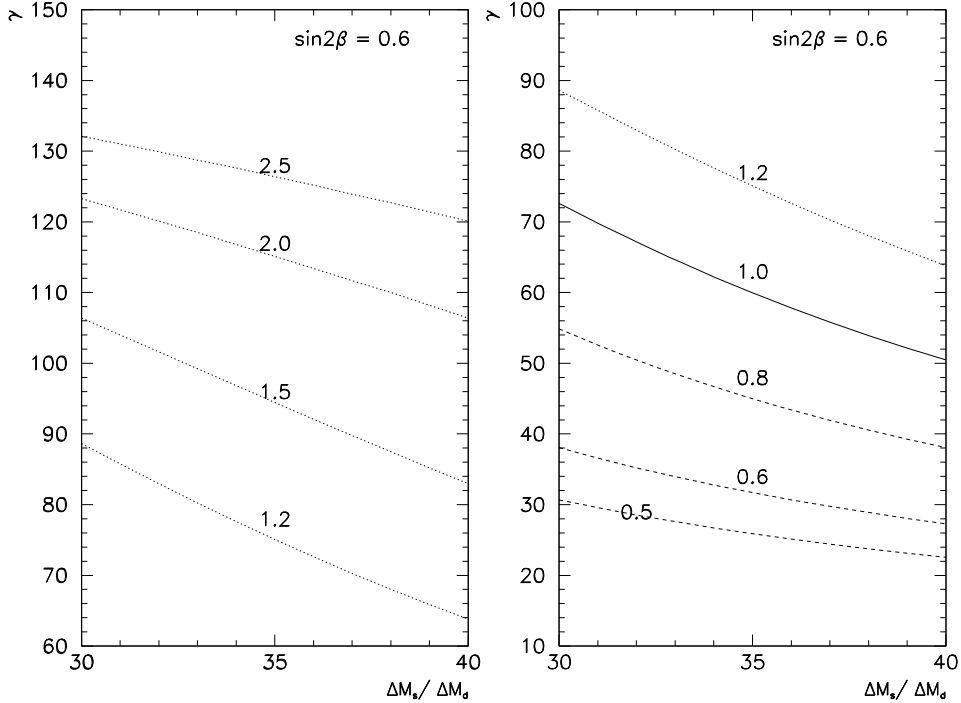}
}
\vspace{0.08in}
\caption[]{
$\gamma$ as a function of $\Delta M_s/\Delta M_d$ for $\sin 2\beta=0.6$ and 
different  $R_{sd}$ \cite{BCRS1,LU}.
\label{Lucja}}
\end{figure}

In order to be sure whether this is indeed the case other theoretically
clean quantities have to be measured. In particular the angle $\gamma$ 
that is more sensitive to new physics contributions than $\beta$.
In this context the measurement of the ratio  $\Delta M_s/\Delta M_d$
will play an important role as for a fixed value of $\sin 2\beta$, 
the extracted value for $\gamma$ is a sensitive function of 
$\Delta M_s/\Delta M_d$ as shown in fig.~\ref{Lucja}. The solid line,
labeled by $R_{sd}=1.0$ in the right plot, represents MFV models. The 
remaining lines, obtained in a general analysis in \cite{BCRS1}, 
represent generalized MFV models in which also significant contributions 
of new operators are possible. See the discussion below (\ref{master}). 
In these models the expression
for $R_t$ in (\ref{Rt}) receives an additional factor $\sqrt{R_{sd}}$. 
For $R_{sd}>1.2$, the angle $\gamma>90^\circ$ is possible provided 
$\Delta M_s/\Delta M_d$ is not too large.

At this point I would like to stress the importance of the precise 
measurements of $a_{\psi K_S}$ and $\Delta M_s/\Delta M_d$ that should 
be available within the coming years. These two measurements taken together 
allow the determination of $\bar\varrho$ and $\bar\eta$ through
\be\label{approx}
\bar\varrho\approx 1-R_t\left[1-\frac{a^2_{\psi K_S}}{8}\right],
\qquad
\bar\eta\approx R_t\frac{a_{\psi K_S}}{2}
\left[1+\frac{a^2_{\psi K_S}}{8}\right]
\ee
with $R_t$ given by (\ref{Rt}). Exact expressions can be found in 
\cite{UUT,BF01,BCRS1}. 
The only theoretical uncertainty in these formulae 
resides in $\xi$ that should be known from lattice calculations 
within a few percent in the next years.
There is another virtue of this particular determination of 
$\bar\varrho$ and $\bar\eta$ that we will discuss in the context of
the last topic on our list. 

What about the angle $\alpha$? For a given $\sin 2\beta$ satisfying 
the CKM unitarity bound \cite{BLO}
\be
\sin 2\beta \le 2 R_b \sqrt{1-R_b^2}
\ee
there are two solutions  for $\alpha$ with $\alpha <90^\circ$ and 
$\alpha>90^\circ$. However, if $\sin 2\beta$ saturates this bound, only
$\alpha=90^\circ$ is possible. An example of a corresponding triangle 
is shown in fig.\,\ref{fig:utdata}. Such a possibility is hinted by a 
large value of $\sin 2\beta$ from Belle and has been advocated by 
Fritzsch and Xing \cite{Harald}
for many years. For $\alpha=90^\circ$ we simply have ($R_t<1$)
\be
\bar\varrho=1-R_t^2, \qquad \bar\eta=R_t\sqrt{1-R_t^2}
\ee
with $R_t$ given by (\ref{RT}) or (\ref{Rt}). Equivalently
\be
\sin\beta=R_b,\qquad \sin\gamma=R_t, \qquad R_b=\sqrt{1-R_t^2}\,.
\ee
Simultaneously the CP asymmetry $a_{\pi^+\pi^-}$ vanishes provided penguin 
pollution (see topic 4) can be neglected. Present BaBar data on
$a_{\pi^+\pi^-}$ are consistent with $\sin 2\alpha=0$.

\subsection{The Ratio \boldmath{$\epe$}}
The ratio $\epe$ measures the relative size of the direct 
($\varepsilon'$) and indirect ($\varepsilon$) 
CP violation in $K_L\to\pi\pi$ decays. One of the highlights of this year 
are the precise measurements of this ratio by NA48 and KTeV collaborations:
\begin{equation}\label{eprime1}
\RE(\varepsilon'/\varepsilon) =\left\{ \begin{array}{ll}
(15.3 \pm 2.6)\cdot 10^{-4} &{\rm (NA48)}~ \cite{NA48}~,\\
(20.7 \pm 2.8)\cdot 10^{-4} & {\rm (KTeV)}~\cite{KTEV}~.
\end{array} \right.
\end{equation}
Combining these results with earlier measurements by
NA31 collaboration at CERN  $ ((23.0\pm 6.5)\cdot 10^{-4})$
 and by the
E731 experiment at Fermilab $ ((7.4\pm 5.9)\cdot 10^{-4})$
gives
the grand average
\be
\RE(\epe) = (17.2 \pm 1.8)\cdot 10^{-4}~.
\label{gae}
\ee
This is another mile stone in CP violation.

On the theoretical side, the short distance contributions to $\epe$
are fully under control \cite{NLOEPE} but the presence of considerable 
long distance 
hadronic uncertainties precludes a precise value of $\epe$ in the SM and 
its extentions at present. 
Consequently while theorists were able to predict the sign and the order 
of magnitude of $\epe$, the range   
\be\label{2}
(\varepsilon'/\varepsilon)_{\rm th} = (5 \ \mbox{to} \
30)\cdot 10^{-4}
\ee
shows that the present status of $(\epe)_{\rm th}$ cannot match the 
experimental one.

It should be emphasized that the short distance contributions to $\epe$ are 
governed by perturbative QCD and electroweak effects, that are very strongly 
enhanced through QCD renormalization group effects active in the range 
$1~\gev\le \mu\le \mt$. Without these effects $\epe$ 
would be a few $10^{-5}$. Consequently the short distance contributions 
determine the order of magnitude of $\epe$.

On the other hand the long distance contributions govern the factor 
``$r$" in 
$\epe= r\cdot 10^{-3}$. These contributions are not yet under control. 
This is clearly seen in an approximate formula \cite{Bosch}
\be
\epe=\IM\lambda_t\cdot F_{\varepsilon'}\qquad (\lambda_t=V^*_{ts} V_{td})
\ee
\be\label{crude}
F_{\varepsilon'}\approx 13\cdot 
\left[\frac{110\mev}{\ms(2~\gev)}\right]^2
\left[\bsi(1-\OEE)-0.4\cdot \bei\left[\frac{\mt}{165\gev}\right]^{2.5}\right]
\left[\frac{\Lms^{(4)}}{340\,\mev}\right]
\ee
where $\bsi$ and $\bei$ represent the hadronic matrix elements of the 
dominant QCD-penguin ($Q_6$) and 
electroweak-penguin ($Q_8$) operators, $\Lms^{(4)}$ is the QCD 
scale and $\OEE$ are isospin breaking effects. The strange quark mass in this 
formula originates in the matrix elements of $Q_6$ and $Q_8$ evaluated in 
the large-N approach. The calculations 
of $\bsi$ and $\bei$ are based on three religions: Large-N approach, Lattice 
approach and the Chiral Quark Model. Large-N approach, formulated for 
weak decays in 1986 by Bardeen, G\'erard and myself \cite{bardeen:87}
and modified in various 
ways by different researchers,  is used by the 
groups in Munich, Dortmund, Granada-Lund, Barcelona-Valencia, 
Beijing and Marseille. The lattice approach in connection with $\epe$ 
is most extensively studied at present in Rome, Southampton,
Brookhaven-Columbia, Geneva-Munich and by CPPACS but the early work of Gupta,
Kilcup and Sharpe should not be forgotten. 
Chiral Quark Model in the context of $\epe$ is the 
domain of the Trieste group. There are 
other small religions in Dubna-Zeuthen, Montpellier and Taipei. The 
results from various groups covering the range in (\ref{2}) are listed in 
table 9 of \cite{Erice}. 
The basic issues in these analyses are
\begin{itemize}
\item
The values of $\bsi$, $\bei$ and $\ms(2\,\gev)$,
\item 
Final State Interactions,
\item 
 Isospin Breaking effects and generally electromagnetic effects.
\end{itemize}
 
These topics are discussed extensively by Donoghue, de Rafael, Martinelli, 
Paschos, 
Colangelo, Golterman and Gardner in these 
proceedings and I will not elaborate on them here. They are also 
reviewed in \cite{Erice,Bosch,REV}.
Instead I would like to investigate \cite{BG01} what is 
the $(\epe)_{\rm exp}$ in 
(\ref{gae}) maybe telling us? To this end let us write 
\be\label{eth}
(\varepsilon'/\varepsilon)_{\rm th} = \mbox{Im} \lambda_t~ [P^{1/2} -
{1\over \omega} P^{3/2}]
\ee
with  $\omega=0.045$ representing the $\Delta I=1/2$ rule.
$P^{1/2}$ is dominated by QCD-penguins, 
in particular the operator $Q_6$. $P^{3/2}$ is governed by isospin breaking 
effects induced by the electric charge difference $\Delta e=e_u-e_d$
(electroweak penguins as $Q_8$) and the mass splitting $\Delta m=m_u-m_d$
represented by $\OEE$ in (\ref{crude}).

\begin{figure}[hbt]
\vspace{0.10in}
\centerline{
\epsfysize=3.0in
\epsffile{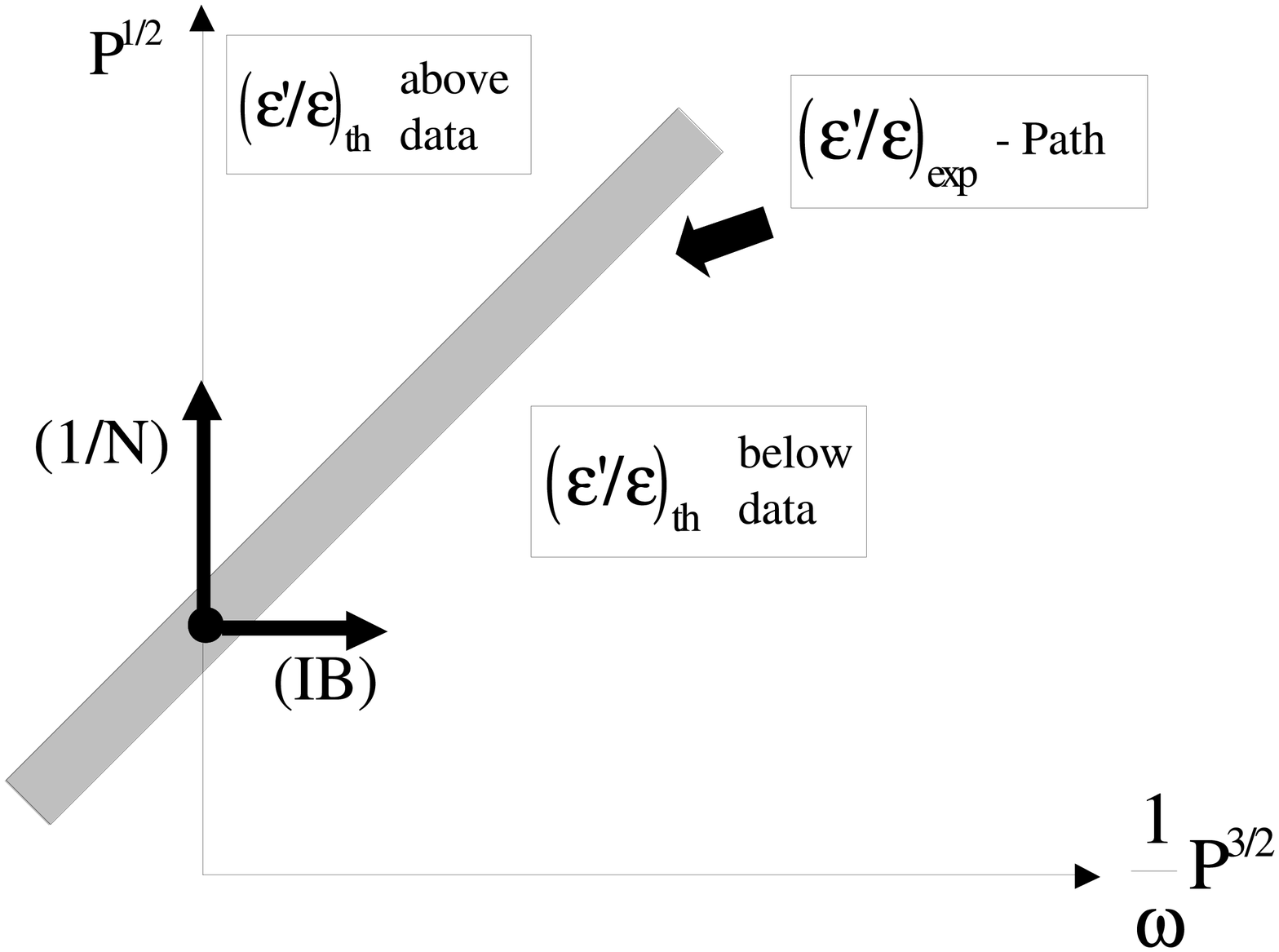}
}
\vspace{0.02in}
\caption{$(\epe)_{\rm exp}$--path in the
$(P^{3/2},P^{1/2})$ plane \cite{BG01}.}
\label{BGPLOT}
\end{figure}

As $\IM\lambda_t$ is known from the analysis of the unitarity triangle,
$(\epe)_{\rm exp}$ in (\ref{gae}) tells us that we are allowed to 
 walk only along a straight path in the $(P^{3/2},P^{1/2})$ plane, 
as illustrated in Fig.\ref{BGPLOT}. 
This path crosses the $P^{1/2}$--axis at 
$(P^{1/2})_0$ = $14.3 \pm 2.8$.

As seen in (\ref{2}), we are still far away from a precise calculation of 
$P^{1/2}$ and $P^{3/2}$. However, we know that isospin-symmetry and 
large-N limit represent two powerful approximations to study
long-distance hadronic physics. So let us ask what we find when we
go to the strict isospin-symmetry limit, setting in particular 
$\alpha_{em}=0$, and take simultaneously the large N limit. As for
$\alpha_{em}\to 0$
electroweak penguins disappear and $\OEE=0$ we land on the  
$P^{1/2}$--axis. On the other hand taking large N limit allows us
to calculate $P^{1/2}$ as in this limit $P^{1/2}$ is given by the $Q_6$ 
penguin with $\bsi=1$ and the smaller $Q_4$ penguin. 
The only uncertainties in $P^{1/2}$ reside now in
$\ms$ and $\Lms^{(4)}$ or equivalently $\alpha_s (M_Z)$.  
To our surprise \cite{BG01}, taking the central values 
$\ms(2 \gev) = 110~{\rm  MeV}$  and $\alpha_s (M_Z) = 0.119$, 
we find $P^{1/2}=14.0$, landing precisely on the $(\epe)_{\rm exp}$-path.
Equivalently
\be\label{BG0}
(\varepsilon'/\varepsilon)_0 = (17.4\pm 0.7) \ 10^{-4}
\ee
where the error results from the error in
$\IM\lambda_t$ obtained with $\hat B_K=3/4$ corresponding to the 
large N limit. 
Clearly, as $\Lms^{(4)}=(340\pm 40)~{\rm MeV}$ 
and $\ms(2 \gev) = (110\pm 20)~{\rm  MeV}$, improvements on these input
parameters are mandatory.

Although this rather intriguing
coincidence between (\ref{gae}) and (\ref{BG0}) seems to indicate small 
$1/N$ and IB corrections, one cannot rule out a somewhat
accidental conspiracy between sizeable corrections
canceling each other that may also include new physics contributions:
\be
{\cal O} (1/N) - {1\over \omega} {\cal O} (IB) \approx 0~.
\ee
The latter equation describes the walking along
the $(\epe)_{\rm exp}$--path.
\subsection{\boldmath{$\kpn$} and \boldmath{$\klpn$}} 
The rare decays $\kpn$ and $\klpn$ are very promising probes of 
flavour physics within the SM and possible extensions, since they are
governed by short distance interactions. They proceed through $Z^0$-penguin
 and box diagrams. As the required hadronic matrix elements can be extracted 
from the leading semileptonic decays and other long distance contributions 
turn out to be negligible \cite{RS}, the relevant branching ratios can be 
computed
to an exceptionally high degree of precision \cite{BB}. The main theoretical
uncertainty in the CP conserving decay $\kpn$ originates in the value of
$\mc(\mu_c)$. It has been reduced through NLO corrections down to $\pm 7\%$
at the level of the branching ratio. The dominantly CP-violating decay 
$\klpn$ \cite{littenberg:89} is even cleaner as only the internal top 
contributions matter. The 
theoretical error for $Br(\klpn)$ amounts to $\pm 2\%$ and is safely 
negligible.

There are three virtues of these decays:
\begin{itemize}
\item
$\imlt$ can be determined directly from $Br(\klpn)$
 \cite{BBSIN}:
\begin{equation}\label{imlta}
\IM\lambda_t=1.36\cdot 10^{-4} 
\left[\frac{170\gev}{\mtb(\mt)}\right]^{1.15}
\left[\frac{Br(\klpn)}{3\cdot 10^{-11}}\right]^{1/2}\,
\end{equation}
without any uncertainty in $\vcb$.
With $\mtb(\mt)$ measured very precisely at Tevatron and later 
at LHC and future linear collider, (\ref{imlta}) offers the cleanest 
method to measure $\imlt$ and effectively the Jarlskog invariant 
$J_{\rm CP}=\imlt (1-\lambda^2/2)\lambda$. 
\item
$\sin 2\beta$ can be determined very cleanly onces both branching ratios 
are known \cite{BBSIN}. Measuring these branching ratios with $10 \%$ 
accuracy allows to
determine $\sin 2\beta$ with an error $\Delta \sin 2\beta=\pm 0.05$. 
Comparision of this determination with the one by means of 
$a_{\psi K_S}(t)$ is particularly well suited for tests of CP violation 
in the SM and offers a powerful tool to probe the physics beyond it 
\cite{BBSIN,NIR}.
\item
The unitarity triangle can be determined very cleanly with the main 
uncertainty residing in the value of the Wolfenstein parameter $A$ or 
equivalently $\vcb$. In particular $\vtd$ can be determined to better than 
$\pm 10\%$.
\end{itemize}

At present we have:
\begin{equation}\label{K+}
Br(\kpn) =\left\{ \begin{array}{ll}
(7.5\pm 2.9) \cdot 10^{-11} & {\rm (SM)} \\
(15^{+34}_{-12})\cdot 10^{-11} &{\rm (E787)}~\cite{E787}
\end{array} \right.
\end{equation}

\begin{equation}\label{KL}
Br(\klpn) =\left\{ \begin{array}{ll}
(2.6\pm 1.2) \cdot 10^{-11} & {\rm (SM)} \\
<5.9\cdot 10^{-7} &{\rm (KTeV)}~\cite{KTeV2}
\end{array} \right.
\end{equation}
where the errors in the SM branching ratios come dominantly from the 
uncertainties in the CKM papameters. The E787 result for $Br(\kpn)$ is 
rather close to the SM expectations, excluding very large non-standard 
contributions. The KTeV result is still four orders of magnitude away from 
the SM prediction for $Br(\klpn)$. The latter branching ratio can be bounded
in a model independent manner using isospin symmetry \cite{NIR}:
\be\label{GN}
Br(\klpn)\le 4.4~Br(\kpn) \le 2\cdot 10^{-9} (90\%~{\rm C.L}).
\ee
The experimental outlook for both decays has been reviewed by Littenberg 
in \cite{LITT00} and at this conference. See also \cite{Bel}. 
We can hope that the efforts by 
experimentalists at Brookhaven, Fermilab and KEK will result in the 
measurements of both branching ratios with $\pm 10\%$ accuracy in the second 
half of this decade. 
\subsection{CP Violation in B Decays}
CP violation in B decays is one of the most important targets of B-factories
and of dedicated B-experiments at hadron colliders. The first results on
$\sin 2\beta$ from BaBar and Belle, discussed already in Section 4.1, are 
very encouraging. These results should be improved over the coming years 
through the new measurements of $a_{\psi K_S}(t)$ by both collaborations 
and by CDF and D0 at Fermilab. An error for $\sin 2\beta$ of $\pm 0.08$
should be achievable by the next summer. Of interest is also the 
measurement of $\sin 2\beta$ through the CP asymmetry in the decay 
$B_d\to\phi K_S$ that proceeds dominantly through penguin diagrams.

In order to search for new physics it is mandatory to measure the 
angles $\alpha$ and $\gamma$ in the UT. Many strategies to measure 
these angles have been proposed in the last decade. 
Reviews can be found in \cite{Erice,BF97,BABAR,LHCB,NEW}. 
Most of these strategies require
simultaneous measurements of several channels in order to remove potential 
hadronic uncertainties present in non-leptonic B-decays. Prime example 
is the measurement of $\alpha$ through the CP asymmetry in 
$B^0_d\to\pi^+\pi^-$,
where the presence of penguin diagrams, in addition to the dominant 
tree diagrams, precludes a clean extraction of $\alpha$ from
$a_{\pi^+\pi^-}(t)$. There are many ideas for determining or eliminating the
penguin component. They are reviewed in \cite{Erice,BF97,BABAR,LHCB,NEW}. 
None of them is straightforward 
and only time will show which of these methods will provide an acceptable
determination of $\alpha$. At present both BaBar and Belle make efforts to
measure $a_{\pi^+\pi^-}(t)$ that gives $(\sin 2\alpha)_{\rm eff}$.
The latter containing penguin contributions does not give the true angle 
$\alpha$. Theorists are then supposed to translate $(\sin 2\alpha)_{\rm eff}$
into the true $\sin 2\alpha$. More about this issue can be found 
in Beneke's talk. It should be emphasized that in view of a 
poor knowledge of $\alpha$ at present (see table\,\ref{TAB2}) even a rough 
measurement of this angle will have an important impact on the UT.

The theoretically cleanest and simultaneously experimentally feasible method
for the determination of the angle $\gamma$ is the full time dependent 
analysis of $B_s\to D_s^+K^-$ and $\bar B_s\to D_s^-K^+$ \cite{adk}. 
This method is
unaffected by penguin contributions but the presence of the expected large 
$B^0_s-\bar B^0_s$ mixing is a challenge for experimentalists. Yet,
LHC-B should be able to measure $\gamma$ in this manner with high 
precision \cite{LHCB}. 
Also $B_c\to D D_s$ could be used for the extraction of
$\gamma$ at the LHC-B \cite{FW01}.

At present the most extensive analyses of the angle $\gamma$ use the four
$B\to \pi K$ channels that have been measured by CLEO, BaBar and Belle.
The main issues here are the final state interactions (FSI), 
SU(3) symmetry
breaking effects and the importance of electroweak penguin
contributions. Several interesting ideas have been put forward
to extract the angle $\gamma$ in spite of large hadronic
uncertainties in $B\to \pi K$ decays 
\cite{FM,GRRO,GPAR1,GPAR3,GPAR2,NRBOUND}.
Reviews can be found in \cite{GPAR3,RFA}.

Three strategies for bounding and determining $\gamma$ have been 
proposed. The ``mixed" strategy \cite{FM} uses 
$B^0_d\to \pi^0 K^\pm$ and $B^\pm\to\pi^\pm K$. The ``charged" strategy
\cite{NRBOUND} involves $B^\pm\to\pi^0 K^\pm,~\pi^\pm K$ and
the ``neutral" strategy \cite{GPAR3} the modes 
$B_d^0\to \pi^\mp K^\pm,~\pi^0K^0$. 
Parametrizations for the 
study of the FSI, SU(3) symmetry
breaking effects and of the electroweak penguin
contributions in these strategies have been presented 
in \cite{GPAR1,GPAR3,GPAR2}. Moreover, general parametrizations by means
of Wick contractions \cite{IWICK,BSWICK} have been proposed. 
They can be used for all two-body B-decays.
These parametrizations should
turn out to be useful when the data improve.

Parallel to these efforts an important progress has been made 
by Beneke, Buchalla, Neubert and  Sachrajda \cite{BBNS1}
through the demonstration that in a large
 large class of non-leptonic two-body
B-meson decays the factorization of the relevant hadronic matrix elements 
follows from QCD in the heavy-quark limit. The resulting factorization 
formula incorporates
elements of the naive factorization approach used in the past but allows to
compute systematically non-factorizable corrections. In this approach
the $\mu$-dependence of hadronic matrix elements is under control.
Moreover spectator quark effects are taken into account and final
state interaction phases can be computed perturbatively. While,
in my opinion, an important progress in evaluating non-leptonic 
amplitudes has been made in \cite{BBNS1}, the usefulness of this approach
at the quantitative level has still to be demonstrated when the
data improve. In particular the role of the $1/\mb$ corrections
has to be considerably better understood. Recent lectures on this
approach can be found in \cite{NEUTASI}.
The techniques developed in \cite{BBNS1} have been used for exclusive 
rare B decays \cite{EXCL}. An interesting proof of factorization for 
$B\to D\pi$ to all orders of $\alpha_s$ has been presented in \cite{BPS}.

There is an alternative perturbative QCD approach to non-leptonic
decays \cite{Li} which has been developed earlier from the QCD 
hard-scattering
approach. Some elements of this approach are present in the
QCD factorization formula of \cite{BBNS1}. The main difference between these
two approaches is the treatment of soft spectator contributions
which are assumed to be negligible in the perturbative QCD approach.
While the QCD factorization approach is  more general and systematic, 
the perturbative QCD approach is an interesting possibility.
Only time will show which of these two frameworks is more successful and
whether they have to be replaced by still more powerful approaches
in the future.

Finally new methods to calculate exclusive hadronic matrix
elements from QCD light-cone sum rules has been developed recently
in \cite{KOD}. This work may shed light on the
importance of $1/\mb$ and soft-gluon effects in the QCD factorization
approach. Reviews of QCD light-cone sum rules can be found in
\cite{LCQCD}.

Returning to phenomenology, as demonstrated in 
\cite{FM,GPAR1,GPAR3,GPAR2,NRBOUND}, 
already CP-averaged $B\to\pi K$ branching ratios
may imply interesting bounds on $\gamma$ 
that may remove a large portion of the allowed range from the analysis
of the unitarity triangle. In particular combining the neutral and
charged strategies \cite{GPAR3} one finds that the existing
data on $B\to \pi K$ favour $\gamma$  in the second quadrant, which is in 
conflict with the standard analysis of the unitarity triangle as
we have seen in section 4.1. Other arguments for $\cos\gamma<0$ using
$B\to PP,~PV$ and $VV$ decays were given in \cite{CLEO99}.
Also the analyses of $B\to\pi K$ in the QCD factorization approach 
\cite{BEBENESA} favour $\gamma>90^\circ$.

In view of sizable theoretical uncertainties in the analyses of
$B\to\pi K$ and of large experimental errors in the
corresponding branching ratios it is not yet clear whether the
discrepancy in question is serious. For instance \cite{CIFRMAPISI}
sizable contributions of the so-called charming penguins to the
$B\to\pi K$ amplitudes could shift $\gamma$ extracted from these
decays below $90^\circ$ but at present these contributions cannot be
calculated reliably. Similar role could be played by annihilation
contributions \cite{Li} and large non-factorizable 
SU(3) breaking effects
\cite{GPAR3}.  Also,  new physics contributions in the electroweak
penguin sector could  shift $\gamma$ to the first quadrant
\cite{GPAR3}.  It should be however emphasized that the problem with
the angle $\gamma$, if it persisted, would put into difficulties not
only the SM but also the full class of MFV models in which the lower
bound on $\Delta M_s/\Delta M_d$ implies $\gamma < 90^\circ$. On the
other hand as seen in fig.~\ref{Lucja} for sufficiently high
values of $R_{sd}$, the angle $\gamma$ resulting from the unitarity
triangle analysis in models containing new operators \cite{BCRS1} 
can easily be in the second quadrant provided
$\Delta M_s/\Delta M_d$ is not too large.
However, this does not happen in the MSSM in the large $\tan\beta$ 
limit, where the presence of new operators results in $R_{sd}<1.0$ 
and in $\gamma$ that is generally smaller than in the SM \cite{BCRS1}.

Another interesting direction is the use of U-spin symmetry.
 New strategies for $\gamma$ using this symmetry have
been proposed in \cite{RF99}. The first strategy involves
the decays $B^0_{d,s}\to \psi K_S$ and 
$B^0_{d,s}\to D^+_{d,s} D^-_{d,s}$.
The second strategy involves $B^0_s\to K^+ K^-$ and $B^0_d\to\pi^+\pi^-$.
These strategies are unaffected by FSI and are only limited
by U-spin breaking effects. They are 
promising for
Run II at FNAL and in particular for LHC-B provided the 
U-spin breaking effects can be estimated reliably \cite{RF99}. 
A method of determining $\gamma$, using $B^+\to K^0\pi^+$ and the
U-spin related processes $B_d^0\to K^+\pi^-$ and $B^0_s\to \pi^+K^-$,
was presented in \cite{GRCW}. A general discussion of U-spin symmetry 
in charmless B decays and more references to this topic can be
found in \cite{G00}.

\subsection{Minimal Flavour Violation Models}
We have defined this class of models in section 3. Here I would like just
to list four interesting properties of these models that are independent 
of particular parameters present in these models. These are:
\begin{itemize}
\item
There exists a universal unitarity triangle (UUT) \cite{UUT} common to all 
these models and the SM that can be constructed by using measurable 
quantities that depend on the CKM parameters but are not polluted by the 
new parameters present in the extensions of the SM. 
The UUT can be constructed, for instance, by using $\sin 2\beta$ from 
$a_{\psi K_S}$ and the ratio $\Delta M_s/\Delta M_d$. 
The relevant formulae can be found in (\ref{approx}) and in
 \cite{UUT,BF01,BCRS1}, where also other 
quantities suitable for the determination of the UUT are discussed.
\item
There exists an absolute lower bound on $\sin 2\beta$ \cite{ABRB} that 
follows from the interplay of $\Delta M_d$ and $\varepsilon_K$.
It depends only on 
$\vcb$ and $\vub$, as well as on the non-perturbative parameters 
$\hat B_K$, $F_{B_d}\sqrt{\hat B_d}$ and $\xi$ entering the standard 
analysis of the unitarity triangle.
A conservative 
scanning of all relevant input parameters gives \cite{Erice}
$(\sin 2\beta)_{\rm min}=0.42$. A less conservative bound of $0.52$ 
has been found in \cite{Perez}.
This bound could be 
considerably improved when the values of $\vcb$, $\vub$, $\hat B_K$, 
$F_{B_d}\sqrt{\hat B_d}$, $\xi$ and -- in particular of 
$\Delta M_s$ -- will be known better \cite{Erice,ABRB}. 
\item
There exists an absolute upper bound on $\sin 2\beta$. It is simply given by 
\cite{BLO}
\be
\label{ubound}
(\sin 2\beta)_{\rm max}=2 R_b^{\rm max}\sqrt{1-(R_b^{\rm max})^2}
\approx 0.82,
\ee
with $R_b$ defined in (\ref{2.94}).
\item
For given $a_{\psi K_{\rm S}}$ and $Br(\kpn)$ only two values of 
$Br(\klpn)$ are possible 
in the full class of MFV models, independently of any new parameters 
present in these models \cite{BF01}. 
Consequently, measuring $Br(\klpn)$ will 
either select one of these two possible values or rule out all MFV models.
Taking the present experimental bound on $Br(\kpn)$ and (\ref{ubound}) 
one finds an absolute upper bound 
$Br(\klpn)<7.1 \cdot 10^{-10}~(90\%~{\rm C.L.})$ 
\cite{BF01} that is stronger than the bound in (\ref{GN}).
\end{itemize}

\section{Twenty Questions}
\setcounter{equation}{0}

{\bf 1. What are the precise values of \boldmath{$V_{ud}$} and 
\boldmath{$V_{us}$} ?}\\
 The unitarity relation $|V_{ud}|^2+|V_{us}|^2+|V_{ub}|^2=1$ is violated
by more than two standard deviations. The only hope is that our understanding
of the errors for $V_{ud}$ and $V_{us}$ is still not what we think. 
Otherwise we have to conclude that some new physics is at work. 
The
improved measurements of these two elements are mandatory.

\par\noindent
{\bf 2.
How accurately can we determine \boldmath{$|V_{ub}|$} 
and \boldmath{$\vcb$} ?}\\
Both determinations are subject to theoretical uncertainties. Recently, 
interesting new methods for the determination of $|V_{ub}|$ from inclusive 
B decays have been proposed \cite{Bauer01}. They could provide, in 
conjunction with the future BaBar and 
Belle measurements, an improved measurement of $R_b$. Both elements 
are very important for the analysis of the unitarity triangle and for 
the predictions of rare decays as the latter are sensitive functions of
the Wolfenstein parameter $A=\vcb/\lambda^2$.

\par\noindent
{\bf 3.
How accurately can we calculate \boldmath{$\hat B_K$, 
$\sqrt{\hat B_d} F_{B_d}$},
\boldmath{$\xi$} and determine \boldmath{$\ms$} and 
\boldmath{$\mc$} ?} \\
All these low energy quantities enter the phenomenology of weak decays 
both in the SM and its extensions and they should be determined with a high 
precision. While ultimately lattice calculations should provide the 
most accurate numbers, QCD sum rule approach will also continue to
be useful for some of these parameters \cite{KOD,LCQCD,Jamin01}.

\par\noindent
{\bf 4.
What is the value of \boldmath{$\Delta M_s$} ? }\\
Possibly we will know it already next summer if CDF and D0 are lucky and 
$\Delta M_s\le 20/ps$ as indicated by LEP analyses. This will be a very 
important measurement, providing accurate values of $R_t$ and $\vtd$.
Simultaneously, combining this measurement with $a_{\psi K_S}$ will 
allow us to determine $\gamma$ in a clean manner as illustrated in 
fig.\,\ref{Lucja}.
Moreover if $\sqrt{\hat B_s} F_{B_s}$ can be calculated accurately by 
lattice or QCD sum rule methods, the measurement of $\Delta M_s$ with
$|V_{ts}|\approx \vcb$ will shed some light on whether MFV is the whole 
story or whether new effective operators have to be taken into account
\cite{BCRS1}.

\par\noindent
{\bf 5.
Is the CKM matrix the only source of flavour and CP violation ?}\\
In order to answer this question the four properties of MFV models listed 
at the end of section 4 will be very useful. 
While last year the lower bound on 
$\sin 2\beta$ from MFV seemed to be a useful quantity for this purpose, 
the new 
BaBar and Belle results are well above this bound and possibly the upper 
bound on  $\sin 2\beta$ in (\ref{ubound}), almost violated by the Belle 
result, could turn out 
to be more interesting in the near future. To this end $R_b$ has to be better 
known. In the long run the last property of the MFV models that involves 
$a_{\psi K_S}$, $\kpn$ and $\klpn$  should be very useful.
There are of course other strategies to answer this question. See Masiero's 
talk and \cite{BCRS1}.
 
\par\noindent
{\bf 6.
What is the optimal error analysis of weak decays ?}\\
There have been already many suggestions in the literature but from my
point of view none of them is fully convincing. It is important to 
make progress here, in particular if new physics contributions will 
turn out to be small.

\par\noindent
{\bf 7.
What are the values of \boldmath{$\bsi$, $\bei$} and \boldmath{$\OEE$}
 and how important are FSI in \boldmath{$\epe$} ?}\\
These are very important questions. Personally, I doubt that these 
questions will be answered very soon but I hope that I am wrong. 
It will be interesting to see what new lattice calculations and new 
analytic efforts can contribute to this issue.

\par\noindent
{\bf 8.
What is the value of \boldmath{$\epe$} in the SM ?}\\
While this question is directly connected with the previous question, 
I pose it here as there are public statements by  
some speakers that $\epe$ in the SM is fully under control, it is in 
perfect agreement with the experiment and the only remaining issue is 
the value of $\ms(2~\gev)$. Such statements are totally misleading. 
While varying the relevant parameters one can certainly fit the 
experimental data, this is not what one wants to do.
Until the question 7 is not answered satisfactorily we do not know  
the precise value of $\epe$ in the SM. My favorite number is still the
one of \cite{AJBLH}. That is $(\epe)_{\rm SM}=7.0 \cdot 10^{-4}$. 

\par\noindent
{\bf 9.
Which are the best decays to look for other signals of direct CP violation 
?}\\
Clearly, in the field of B-decays, one should look at CP asymmetries 
in $B^\pm$ decays. Until now no effects have been found. There are also
efforts by NA48 to measure direct CP violation in $K^\pm$ decays. 
Unfortunately none of these decays, similarly to $\epe$, is theoretically 
clean. In this context one should ask the next question. 

\par\noindent
{\bf 10.
Who will give more money for \boldmath{$\kpn$} and in particular for 
\boldmath{$\klpn$}?} \\
The nonvanishing branching ratio $Br(\klpn)$ higher than $10^{-13}$ is
a signal of CP violation in the interference of mixing and decay. But
at the level of $3 \cdot 10^{-11}$, as expected in the SM, it is a 
clear signal of CP violation in the decay amplitude or equivalently 
of direct CP violation. Higher branching ratios are still possible 
beyond the SM \cite{BS00}. As $Br(\klpn)$  in the SM and in its extensions 
is free of hadronic uncertainties it is exceptional in the field of 
weak decays and it would be a crime if one did not measure it. Similar 
comments apply to $\kpn$. I am convinced that once both branching ratios
have been measured to better than $10\%$ accuracy our understanding 
of flavour violation and CP violation will improve considerably, 
independently of other measurements performed in this decade that
in most cases suffer from substantially larger theoretical uncertainties
than these two golden decays.

\par\noindent
{\bf 11.
How large are CP-conserving and indirectly CP-violating contributions to   
\boldmath{$\kpe$} ?}\\
This decay is believed to be dominated by the contribution from 
direct CP violation that can be calculated very reliably and is 
 expected to give 
$Br(\kpe)_{\rm dir}=(4.3\pm 2.1)\cdot 10^{-12}$ where the error is 
dominated by the CKM uncertainties. In order 
to be able to compare this result with the future data both the 
CP conserving contribution (estimated to be well below $2\cdot 10^{-12}$) 
and the indirectly CP-violating contribution (to be determined by KLOE 
at Frascati) have to be known. The most recent experimental bound from 
KTeV reads $Br(\kpe)\le 5.1\cdot 10^{-10} (90\% {\rm C.L.)}$ leaving 
considerable 
room for new physics contributions. A nice summary of the 
theoretical situation with the relevant references is given by D' Ambrosio 
and Isidori in \cite{Bel}.

\par\noindent
{\bf 12.
Can we ever extract the short distance component of \boldmath{$\kmm$}?}\\
The absorptive part to this decay, determined from 
$K_L\to\gamma\gamma$, is very close to the experimental branching 
ratio : $(7.18\pm 0.17)\cdot 10^{-9}$ from E871 at Brookhaven. In order 
to extract the 
short distance dispersive contribution (estimated to give 
$(0.9\pm 0.3)\cdot 10^{-9}$ in the SM), the long distance dispersive 
contribution has to be known. There are different opinions 
whether the latter contribution can ever be reliably computed \cite{KLMUMU}.
In the positive case one would get a useful measurement of the 
parameter $\bar\varrho$.

\par\noindent
{\bf 13.
How precisely can we determine \boldmath{$\alpha$} and 
\boldmath{$\gamma$} from B-decays 
before LHC-B and BTeV ?}\\
I have addressed this issue already in section 4.4. The answer to
this question depends on the answer to the next question.

\par\noindent
{\bf 14.
How can we get non-leptonic two-body B-decays fully under control ?}\\
Clearly, there has been a considerable progress in calculating 
branching ratios for these decays in the last two years. However, 
from my point of view the situation is far from satisfactory and 
I expect that it will take a considerable amount of efforts by 
experimentalists and theorists before the dynamics of these decays 
will be fully understood.

\par\noindent
{\bf 15.
Is the angle \boldmath{$\gamma$} extracted from 
\boldmath{$B\to\pi K$} decays consistent 
with the UT fits?}\\
This is an important question that requires some progress on the last 
question.

\par\noindent
{\bf 16.
What are the prospects for precise measurements of 
\boldmath{$B\to X_{s,d} \gamma$}, 
\boldmath{$B\to X_{s,d}\mu\bar\mu$} and \boldmath{$B\to X_{s,d}\nu\bar\nu$}, 
for coresponding exclusive channels, \boldmath{$B_{s,d}\to \mu\bar\mu$}
 and related 
theory ?}\\
This is clearly an exciting field with new interesting theoretical papers on 
QCD factorization in exclusive decays $B\to K^*(\varrho)\gamma$ and $B\to K^*\mu\bar\mu$ 
\cite{EXCL} and new very 
relevant analyses of $B_s\to \mu\bar\mu$ at large $\tan\beta$ in 
supersymmetry \cite{BMUMU}. 
The coming years should be very exciting for this field in view of the
new data from BaBar, Belle, CDF and D0.

\par\noindent
{\bf 17.
Do we see any new physics in charm and hyperon decays ?}\\
It is still to early to claim anything of this sort but these 
decays could be the first to provide some hints for new physics 
in spite of the fact that they are not theoretically clean. 
See the talks by Bigi and He in these proceedings \cite{BiHe}.

\par\noindent
{\bf 18.
What are the lowest values of electric dipole moments still compatible 
with low energy supersymmetry ?} \\
It will be exciting to follow the new experimental progress in this field 
and to see how far one can adjust various supersymmetric parameters in
case no signal is found soon. As non-vanishing electric dipole moments
signal CP violation in flavour diagonal transitions, that are very 
strongly suppressed in the SM, their observation will certainly signal 
the presence of new CP-violating phases and might help to explain the 
origin of matter-antimatter asymmetry in the universe.

\par\noindent
{\bf 19.
What are the prospects for leptonic flavour and CP violation ?}\\
This field is experiencing a great push due to the discovery of 
neutrino oscillations. Every day several papers appear. Progress both 
in theory and experiment is to be expected in the coming years.

\par\noindent
{\bf 20.
What is the (indirect) impact of \boldmath{$(g-2)_{\mu}$}
 on weak decays?}\\
The possible discrepancy between the Brookhaven measurement \cite{Brookhaven}
and the SM \cite{CZARMAR} is clearly one of the highlights of this year 
but I think we should not get overexcited in view of considerable theoretical 
uncertainties. 
New improved data as well as theoretical efforts 
will hopefully make the situation clearer. There is an avalanche
of papers in this field, in particular in the framework of supersymmetry.
In the context of this question, an interesting relation
between $(g-2)_{\mu}$ and $B_s\to \mu\bar\mu$ 
at large $\tan\beta$ in supersymmetry has been pointed out in \cite{DDN}.

{\bf Acknowledgements}

I would like to thank the organizers for inviting me to such an 
interesting symposium and Frank Kr\"uger for invaluable comments on
the manuscript.
Thanks are also due to the Max-Planck-Institute (Werner Heisenberg Institute 
for Physics) in Munich for financial support related to this symposium.
The work presented here has been supported in part by the German 
Bundesministerium f\"ur
Bildung und Forschung under the contract 05HT1WOA3 and the 
DFG Project Bu. 706/1-1.

\vfill\eject


\begin{thebibliography}{80}
\bibitem{LU}
I thank {\L}ucja S{\l}awianowska for providing this plot. The results for 
different $\sin 2\beta$ can be found in \cite{BCRS1}.
\end{thebibliography}
\end{document}